# GEANT4 MODELING OF ENERGY SPECTRUM OF FAST NEUTRONS SOURCE FOR THE DEVELOPMENT OF RESEARCH TECHNIQUE OF HEAVY SCINTILLATORS

Viktoriia Lisovska[1], Tetiana Malykhina[1]*,
Valentina Shpagina[2], Ruslan Timchenko[1]
[1]*Kharkiv V.N. Karazin National University*
*4, Svobody sq., 61022, Kharkiv, Ukraine*
[2]*National Science Center "Kharkiv Institute of Physics and Technology"*
*1, Akademichna str., 61108, Kharkiv, Ukraine*
*E-mail: malykhina@karazin.ua

The proposed work demonstrates the results of creating and investigating the mathematical model of the source of fast neutrons. Computer modeling of the energy spectrum of fast neutrons was carried out for $^{239}$PuBe neutron source. The model of the source of fast neutrons has been developed. Neutrons in this model have an energy spectrum from 100 keV to 11 MeV with 100 keV step. Simulation is performed by the Monte-Carlo method. The model carrier is a computer program developed in the C++ programming language in the Linux operating system environment, using the Geant4 toolkit.
All necessary classes describing low-energy models were used for the simulation of the passage of neutrons through materials of detectors. Those take into account the elastic scattering, inelastic scattering, radiative capture and fission. We consider these processes because models of processes implemented in our software will be also used for other problems of neutrons transport, for example, for passing neutrons through various substances, and for conducting virtual laboratory works. The PhysicsList class of our program contains classes G4NeutronHPElastic, G4NeutronHPElasticData, G4NeutronHPInelastic, G4NeutronHPInelasticData, G4NeutronHPCapture, G4NeutronHPCaptureData, etc. based on the NeutronHP model for neutron interactions at low energy, as well as the neutron data library G4NDL4.5.
Diagrams containing energy spectra of a source of fast neutrons modeled in two ways are presented in the paper. The analysis of the obtained energy spectra is carried out. Virtual nuclear physics experiments are carried out with the aim of testing the elaborated neutron-matter interaction model. The processes occurring in scintillator substances during the passage of fast neutrons through them, have been studied. $10^8$ neutrons were used as primary particles emitted isotropically, and we used our simulation results of $^{239}$PuBe neutron source to describe the initial energy spectrum.
The created model of $^{239}$PuBe neutron source can be used for the investigation of scintillation detectors $Bi_4Ge_3O_{12}$, $CdWO_4$, $Gd_2SiO_5$ and others, as well as studying their characteristics. Processes in heavy oxide scintillators substance during the registration of fast neutrons can be studied using the developed model. It is shown that for registration of the flow of neutrons from $^{239}$PuBe neutron source, using $Bi_4Ge_3O_{12}$ or $CdWO_4$ scintillators is more preferable. Results of the virtual nuclear physical experiments satisfy the published experimental data.

With the aim of environmental monitoring and control of radiation hazard facilities, the devices for neutrons detection are developed by leading research centers [1]. Various classic methods are applied for registration of neutrons depending on neutrons energy. If neutrons energies exceed 10 MeV, its registration is based on the usage of compound with carbon and the study of interactions of neutrons with carbon nuclei [2]. Registration of neutrons with energies from 100 keV to 10 MeV is performed by scattering them on hydrogen-containing substances, and consequent registration of recoil protons.

The possibilities of detection of fast neutrons flux by heavy inorganic scintillators are investigated by scientists of V.N. Karazin Kharkiv National University in collaboration with scientists from Kharkiv Institute of Scintillation Materials [1].

In developing of devices for radiation detection, the usage of mathematical modeling allows to conduct a model experiment, and investigate the characteristics of developed detectors. The laboratory detector's experiments are carried out in harmful and dangerous working conditions due to the ionizing radiation. Computer simulation allows us to evaluate some technical parameters of the device being developed. Therefore, computer simulation is an important stage of detectors developing.

## FORMULATION OF THE PROBLEM

The possibility of practical usage of heavy inorganic scintillators for neutron detection is investigated in the work [1]. However, for practical use of detectors it is necessary to estimate contributions of various mechanisms and processes that occur in the matter of these scintillators. This task can be solved using mathematical modeling and analysing its results.

The laboratory experiments were carried out with $^{239}$PuBe neutron source; therefore an important phase of investigations is the development of mathematical model of the neutron radiation source. The goal of our work is computer simulation of the energy spectrum of $^{239}$PuBe neutron source that will be used in programs for simulation of the passage of fast neutrons through matter.

There are published articles [3, 4] where one can see the experimental energy spectra of $^{239}$PuBe neutron sources. However, these data have the too wide energy step and therefore can cause significant systematic errors in virtual experiment, which use them without preliminary processing.

### DEVELOPING OF THE MODEL FOR FAST NEUTRONS TRANSPORT

For study of the processes in components of the experimental facility used for fast neutrons detection, the mathematical modeling of fast neutrons passage through matter was carried out. The modeling was carried out using the Monte Carlo method.

The mathematical model is a computer program developed by our team for the simulation of passage of neutrons from $^{239}$PuBe neutron source through various substances. The program developed in C++ language with using Geant4 toolkit [5]. Description of the model includes physical and chemical properties of the materials of various facility components, its location and relative position as well as neutron source parameters. The special module contains description of neutron source. For developing of neutron source model the neutron source energy spectra from articles [3, 4] were digitized. Spectra in these works have the form of a histogram with wide (data from [3]) and non-uniform (data from [4]) step of energy.

For the purpose of obtaining a uniform energy step, the data was interpolated by Lagrange polynomials [6] after digitizing. Lagrange polynomial interpolation is a convenient method for such problems. The interpolation was carried out consistently with quantity of nodes from 5 to 7. The fragmentation into smaller segments is necessary to avoid accumulation of errors in the process of interpolation [6]. The obtained data was used as input for G4PrimaryGenerator class [7], which is mandatory in each computer program based on Geant4 toolkit.

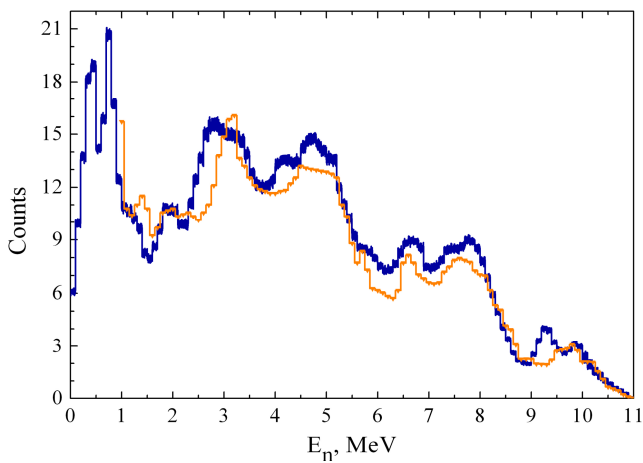

Fig. 1. The result of the G4PrimaryGenerator module: simulated neutron source spectra obtained from data [3] (light line in the chart) and [4] (dark line in the chart).

Figure 1 illustrates neutron source energy spectra modeled from two data sets [3, 4]. Energy spectra in Figure 1 were obtained as a result of the implementation of the G4PrimaryGenerator module as well as the G4GeneralParticleSource module, which uses two data sets [3, 4] as input data.

The G4GeneralParticleSource class [7] was applied for high-energy neutrons transport. This class allows specifying the spectral, spatial and angular distributions of primary neutrons. The flux of $10^7$ primary neutrons was simulated with interpolated energy spectra in order to obtain a suitable statistics. Neutrons were radiated isotropically in this model experiment.

Energy spectra of primary neutrons (Fig. 1) are results of work of G4PrimaryGenerator module, and were written in steps of 100 keV for each sets of initial data, obtained as a result of interpolation [3, 4]. Thus, obtained data is given to a form convenient for further use by modules of our programs.

It should be noted that spectrum of primary neutrons based on data from work [4] is more preferable for further calculations because it contains energy data of primary neutrons in the energy range from 100 keV to 11 MeV. Unlike this energy spectrum, the spectrum based on [3] does not contain data from 100 to 1000 keV.

Our program has a user-friendly interactive mode. We have provided a number of keywords in the program, using which one can "turn on" or "turn off" some processes (previously specified in our program). For example, the command `/process/inactivate/inelastic` entered by the user in an interactive mode allows one "turns off" from consideration the processes of inelastic scattering.

### BRIEF DESCRIPTION OF PHYSICAL PROCESS MODELS AND CLASSES
### FOR LOW ENERGY NEUTRON TRANSPORT USED IN OUR PROGRAM

The high-precision low energy neutron transport is implemented in corresponding classes of the Geant4 toolkit, namely, G4NeutronHPElastic, G4NeutronHPCapture, G4NeutronHPFission, G4NeutronHPInelastic. According to [7], the cross section data for low energy neutron transport are organized in a set of files that are read in by the corresponding data set classes. The classes accessing the total cross section of the individual processes, i.e., the cross section data for high-precision low energy neutron transport, are G4NeutronHPElasticData, G4NeutronHPCaptureData, G4NeutronHPFissionData, and G4NeutronHPInelasticData [8].

The NeutronHP package in Geant4 describes high-precision neutron interactions [8] and contains information about cross sections, angular distribution of the emitted particles, energy spectra of the emitted particles, number of neutrons per fission, fission product yields etc., this data is placed in G4NDL4.5 files which represent Geant4 Nuclear Data Files. A detailed description of the models of these physical processes is given in [7, 8].

## SERIES OF VIRTUAL EXPERIMENTS

In further model experiments, the opportunity of signal recording of $^{239}$PuBe neutron source from distance 1000 mm was investigated. Initially, in real-life experiments the lead shielding was used for protection against gamma rays emitted from the $^{239}$PuBe neutron source (besides neutrons). A center of this lead screen was situated at half distance to the detector, and had thickness 50 mm. The transverse size of lead shielding was 65 mm × 65 mm.

Figure 2 schematically demonstrates the arrangement scheme of facility elements. The proportions of the detector are 20 mm × 20 mm × 20 mm.

The first series of the virtual experiment simulated the conditions in which these practical experiments were carried out. The main aim of this virtual experiment is studying how the initial spectrum changes after passage through the lead shielding. The resulting neutrons spectrum at the distance of 1000 mm from $^{239}$PuBe neutron source, as well as the initial spectrum, are shown in Figure 3.

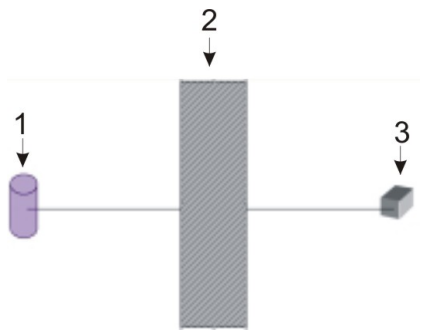
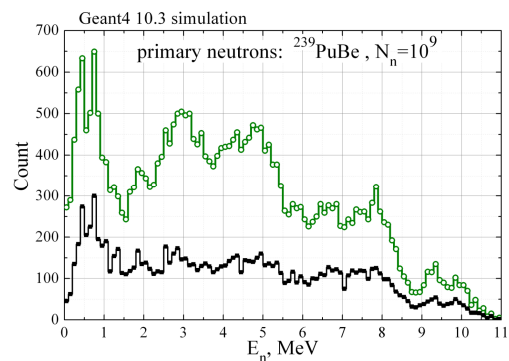

Fig.2. Arrangement scheme of facility elements in the first series of virtual experiments:
1 is neutron source; 2 is protective screen; 3 is detector

Fig. 3. Comparison of the initial neutrons spectrum (with using a protective Pb screen) and the result neutrons spectrum. Light line with empty circles is the initial spectrum; straight line is the result spectrum at the distance of 1000 mm from $^{239}$PuBe neutron source

It can be seen (Fig. 3) that the general view of neutrons spectrum has been changed in energy range from 2 MeV to 5 MeV, however the quantity of neutrons before the detector (at the distance of 1000 mm from the source) significantly decreased. As a result of analysis of the virtual experiment for $N_n=10^9$ primary neutrons emitted from $^{239}$PuBe neutron source, 31763 neutrons were observed before the detector without Pb shielding, and 11825 neutrons in case of using Pb shielding, this difference is explained by inelastic scattering.

The next series of virtual experiments was necessary for the development of a methodology for conducting real experiments, in which the number of neutrons detected at a distance of 1000 mm must be constant. Figure 4 schematically demonstrates the arrangement scheme of facility elements in the second series of virtual experiments. $^{239}$PuBe neutron source is placed into a lead sphere. The radius of the sphere is 52 mm. The distance from the neutron source to the detector is 1000 mm.

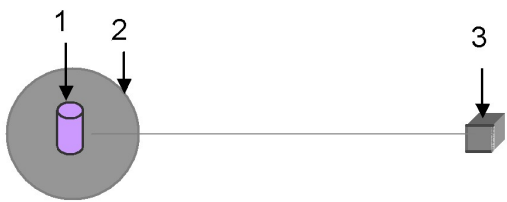
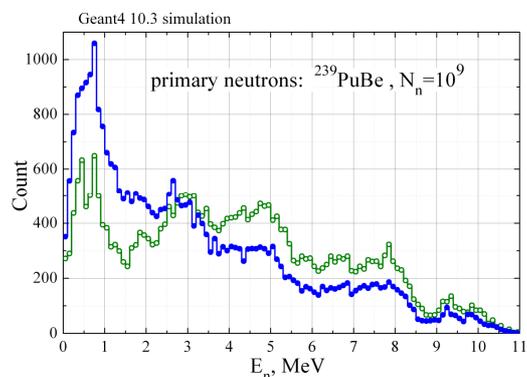

Fig.4. Arrangement scheme of facility elements in the second series of virtual experiments:
1 is neutron source; 2 is a lead sphere; 3 is detector

Fig. 5. Comparison of the initial neutrons spectrum ($^{239}$PuBe source is into Pb sphere) and the result neutrons spectrum. Light line with empty circles is the initial spectrum; dark straight line is the result spectrum at the distance of 1000 mm from $^{239}$PuBe neutron source

It can be seen (Fig. 5) that the general view of the result neutrons spectrum has been changed significantly in the energy range from 0.5 MeV to 5 MeV, especially in the energy range 0.5-2 MeV. As a result of analysis of the second series of virtual experiments for $N_n=10^9$ primary neutrons emitted isotropically from $^{239}$PuBe neutron source, 31763 neutrons were observed before the detector without the lead sphere, and 31694 neutrons in case of using the lead sphere surrounding the neutrons source.

## ESTIMATION OF SECONDARY GAMMA QUANTA QUANTITY

The energy spectra of all secondary gamma quanta arising from the model experiment are shown in Fig. 6 (a).

Analyzing the simulation data, we found out that only negligible number of gamma quanta were registered by the LiI detector. It cannot affect the quality of detection of fast neutron flux. Figure 6 (b) illustrates the energy spectra of gamma quanta incoming in the detector.

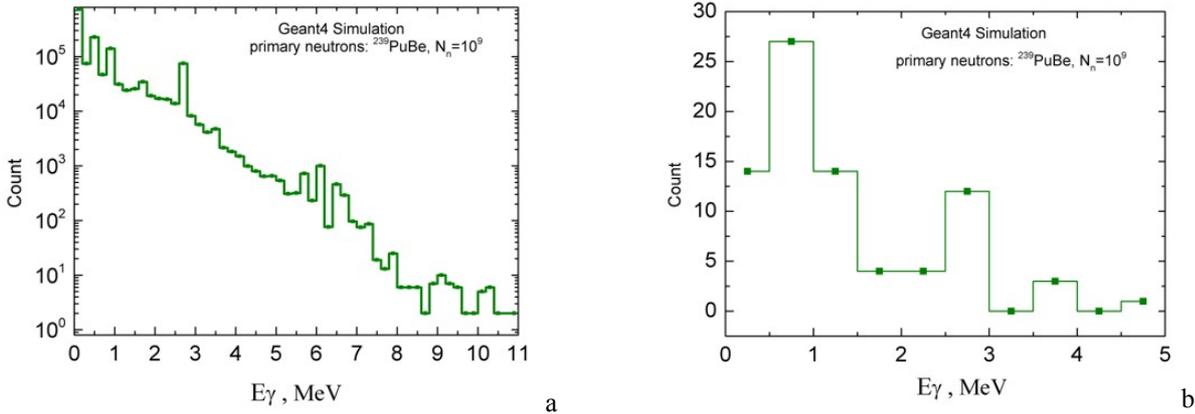

Fig. 6. Spectra of modeled secondary gamma quanta:
(a) is spectrum for all gamma quanta; (b) is spectrum for incoming in detector gamma quanta

The total number of secondary gamma quanta produced in the process of model experiment is about $1.5\times10^7$, while the number of gamma quanta registered by detector is 79.

The further laboratory tests confirmed the opportunity of applying heavy inorganic scintillators of small size for fast neutron flux registration from $^{239}$PuBe neutron source.

## COMPARISON OF RESULTS WITH PUBLISHED DATA

Experimental measurements and analysis of detection efficiency were carried out for fast neutrons from $^{239}$PuBe neutron source using oxide scintillators: $Bi_4Ge_3O_{12}$, $CdWO_4$, $Gd_2SiO_5$ as well as CsI(Tl), NaI(Tl), LiI(Eu) [9]. Fast neutrons registration efficiencies obtained experimentally [9] by heavy inorganic oxide scintillators (Z>50) that have the same size, reach values from 42 to 48 percent and presented in Table 1. Results of measurements of the neutron fluxes registration efficiency by different scintillators in the equivalent energy range for electrons (gamma-quanta) of 20-300 keV [9] are presented in the right column of the Table 1.

Inelastic scattering reaction (n, n'γ) is a one of reliable mechanisms providing a high efficiency of fast neutron detection [7] by oxide scintillators. Figure 7 demonstrates modeled values of deposited energy in detectors.

Table 1.
Efficiency of fast neutron registration, % [9]

| Scintillator (monocrystalline solid) | $Z_{eff}$ | Efficiency of fast neutron registration, % reaction (n, n'γ) |
|---|---|---|
| BGO | 75 | 48 |
| GSO | 59 | 46 |
| CWO | 66 | 42 |
| CsI(Tl) | 54 | 20 |
| NaI(Tl) | 51 | 18 |
| $^6$LiI(Eu) | 52 | 25 |

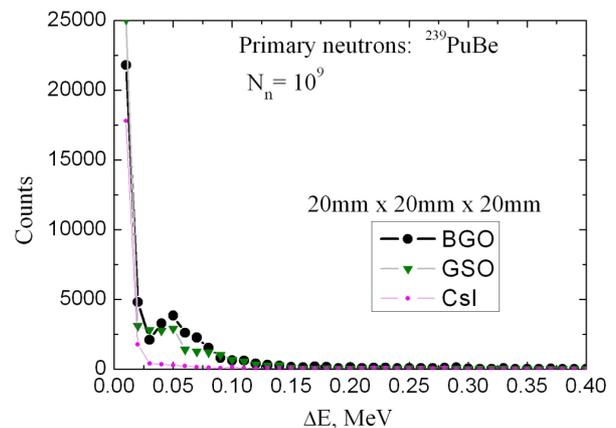

Fig. 7. Modelled values of absorbed energy in detectors for BGO, GSO, and CsI.

Small peaks are visible at ΔE≈0.05 MeV for BGO and GSO scintillators. As a result of this analysis, the maximum peak value is 3830, therefore the statistical uncertainty of the Monte Carlo method is $1/\sqrt{3830} \approx 0.016$. This statistical uncertainty is acceptable for preliminary and estimational calculations.

Using preliminary virtual experiments, one can study mechanisms and reactions occurring in matter of scintillators during the registration of fast neutrons.

## CONCLUSION

As a result of this work, the characteristics of the neutron source have been modeled and used for the study of response of scintillator detectors to neutron flux of $^{239}$PuBe neutron source.

The convenient form of representation of fast neutrons energy spectrum from $^{239}$PuBe neutron source has been obtained for use in computer programs for analyzing the characteristics of scintillators.

It is shown that the model based on data [4] is more preferable for further usage, because it contains neutrons with energies below 1 MeV in the initial spectrum. This is significant for neutron flux detection by heavy inorganic scintillators ($Bi_4Ge_3O_{12}$, $CdWO_4$, $Gd_2SiO_5$ and others).

It is proposed to improve the methodology for conducting experiments for studying the response of heavy oxide scintillators by providing preliminary virtual experiments.

Based on conducted research, it can be concluded that heavy oxide scintillators, which at the same time are efficient gamma-detectors [9], allow to create highly efficient gamma-neutron detectors that provide high efficiency of fissile radioactive materials detection.

**ORCID IDs**
**Viktoriia Lisovska** https://orcid.org/0000-0003-1237-7959, **Tetiana Malykhina** https://orcid.org/0000-0003-0035-2367,
**Valentina Shpagina** https://orcid.org/0000-0002-6202-7474, **Ruslan Timchenko** https://orcid.org/0000-0003-4983-9168